\documentclass[12pt,oneside]{article}
\usepackage{amsfonts,amssymb,graphicx}
\usepackage{amsmath}
\usepackage{subfigure}
\def\no{\noindent}
\def\be{\begin{equation}}
\def\ee{\end{equation}}
\def\bea{\begin{eqnarray}}
\def\eea{\end{eqnarray}}

\def\<{\langle}
\def\>{\rangle}
\def\~{\tilde}
\def\s{\sigma}

\def\a{\alpha}

\def\arctanh{{\rm arctanh \,}}

\def\tm{\widetilde{m}}

\begin{document}
%


\begin{center}
\vspace{1truecm}
{\bf\sc\Large parameter evaluation of a simple mean-field model \\ of  social interaction }\\
\vspace{1cm}
{Ignacio Gallo$^\dagger$, \quad Adriano Barra$^{*\dagger}$, \quad Pierluigi Contucci$^\dagger$}\\
\vspace{.5cm}
{\small $^\dagger$Dipartimento di Matematica, Universit\`a di Bologna,\\ {e-mail: {\em gallo@dm.unibo.it, contucci@dm.unibo.it}}} \\
{\small $^*$Dipartimento di Fisica, Sapienza Universit\`a di Roma,\\ {e-mail: {\em adriano.barra@roma1.infn.it}}\\}
\end{center}
\vskip 1truecm
\begin{abstract}\noindent
The aim of this work is to implement a statistical mechanics theory of social interaction, generalizing econometric
discrete choice models. A class of simple mean field discrete models is introduced and discussed both from the theoretical and phenomenological point of view. We propose a parameter evaluation procedure
and test it by fitting the model against three families of data coming from different cases: the estimated interaction
parameters are found to have similar positive values, giving a quantitative
confirmation of the peer imitation behaviour found in social psychology. Moreover all the values of the interaction
parameters belong to the phase transition regime suggesting its possible role in the study of social systems.
\end{abstract}

\section{Introduction}

In recent years there has been an increasing awareness towards the
problem of finding a quantitative way to study the role played by
human interactions in shaping behaviour observed at a population
level. Indeed, as early as in the nineteen-seventies the dramatic
consequences of including peer interaction in a mathematical model
have been recognized independently by the physics \cite{follmer} (and more recently \cite{galam}),
economics \cite{schelling} and sociology \cite{granovetter}
communities. The conclusion reached by all these studies is that
mathematical models have the potential of describing several features
of social behaviour, among which, for example, the sudden
shifts often observed in society's aggregate behaviour \cite{kuran}, and that
these are unavoidably linked to the way individual people
influence each other when deciding how to behave.

The possibility of using such models as a tool of empirical
investigation, however, is not found in the scientific literature
until the beginning of the present decade \cite{durlauf}: the
reason for this is to be found  in the intrinsic difficulty
of establishing a methodology of systematic measurement for social
features. Confidence
that such an aim might be an achievable one has been boosted by
the wide consensus gained by econometrics following the Nobel
prize awarded in $2000$ to economist Daniel Mcfadden for his work
on probabilistic models of discrete choice, and by the increasing
interest of policy makers for tools enabling them to cope with the
global dimension of today's society \cite{halpern, gacocoga}.

This has led very recently to a number of studies confronting directly the challenge of measuring numerically
social interaction for {\it bottom-up} models, that is models deriving macroscopic phenomena from assumptions
about human behaviour at an individual level \cite{bouchaud1, salganik, bouchaud2, soet}.

These works show an interesting interplay of methods coming from
econometrics \cite{greene}, statistical physics \cite{ellis} and
game theory \cite{nesh}, which reveals a substantial overlap in
the basic assumptions driving these three disciplines. It must also
be noted that all of these studies rely on a simplifying assumption
which considers interaction working on a global uniform scale,
that is on a {\it mean field} approach. This is due to the inability, stated in \cite{watts}, of
existing methods to measure social network topological structure in any detail.
It is expected that it is only matter of time before technology allows to overcome this difficulty: one of the roles
of today's empirical studies is then to assess how much information can be derived from the existing kind of data
such as that coming from surveys, polls and censuses.

This paper considers a mean field model that highlights the possibility of using the methods of discrete choice
analysis to apply a statistical mechanical generalization of the model introduced in \cite{durlauf}. The aim
of the paper is two-fold. On one hand we are interested in assessing how well the simplest instance of such a model
fares when confronted with data, and on the other, we'd like to propose a simple procedure of estimation, based
on a method developed by Berkson \cite{berkson}, that we feel might be very appealing for models at an early
stage of development.

\section{The model}

Consider a population of individuals facing with a ``YES/NO''
question, such as choosing between marrying through a religious or
a civil ritual, or voting in favor or against of death penalty in
a referendum. We index individuals by $i, \ i=1...N$, and assign a
numerical value to each individual's choice $\s_i$ in the following
way:
$$
\s_i = \left\{ \begin{array}{lll}
     +1 \textrm{ if $i$ says YES}
    \\
     -1 \textrm{ if $i$ says NO}
    \end{array}, \right.
$$
Consistently with the widespread use of logit models in econometrics \cite{luce}, and with the
statistical mechanics approach to modelling systems of many interacting agents \cite{durlauf, chandler},
we assume that the joint probability distribution of these choices may be well approximated by 
a Boltzmann-Gibbs distribution corresponding to the following Hamiltonian
$$
H_N(\s)=-\sum_{i,l=1}^N J_{il}\s_i\s_l - \sum_{i=1}^N h_i \s_i.
$$

Heuristically, this distribution favours the agreement of the
people's choices $\s_i$ with some external influence $h_i$ which
varies from person to person, and at the same time favours
agreement of a couple of people whenever their interaction
coefficient $J_{il}$ is positive, whereas favors disagreement
whenever $J_{il}$ is negative.

Given the setting, the model consists of two basic steps:
\begin{itemize}
    \item[1)] A parametrization of quantities $J_{il}$ and of $h_i$,
    \item[2)] A systematic procedure allowing us to ``measure'' the parameters characterizing the model, starting
              from statistical data (such as surveys, polls, etc).
\end{itemize}

The parametrization must be chosen to fit as well as possible the
data format available, in order to define a model which is able to make
good use of the increasing wealth of data available through
information technologies.

\section{Discrete choice}

Let us first consider our model when it ignores interactions
$J_{il}\; \equiv \; 0 \; \forall \; i,l \; \in \; (1,...,N)$, that is
$$
H_N(\s)= - \sum_{i=1}^N h_i \s_i.
$$

The model shall be applied to data coming from surveys, polls, and censuses, which means that
together with the answer to our binary question, we shall have access to information characterizing
individuals from a socio-economical point of view.
We can formalize such further information by assigning to each person a vector of \emph{socio-economic attributes}
$$
a_i=\{a_i^{(1)}, a_i^{(2)},..., a_i^{(k)}\}
$$
where
$$
a_i^{(1)} = \left\{ \begin{array}{l}
    1 \textrm{ for $i$ Male}
    \\
    0 \textrm{ for $i$ Female}
    \end{array}, \right.
$$
or
$$
a_i^{(2)} = \left\{ \begin{array}{l}
    1 \textrm{ for $i$ Employee}
    \\
    0 \textrm{ for $i$ Self-employed}
    \end{array}, \right.
$$
etc.

We choose to exploit the supplementary data by assuming that
$h_i$ (which is the ``field'' influencing the choice of $i$) is a function
of the vector of attributes $a_i$. Since for the sake of simplicity we chose
our attributes to be binary variables, the most general form for $h_i$ turns
out to be linear
$$
h_i=\sum_{j=1}^k \a_j a_i^{(j)}+\a_0
$$
so that the model's parameters are given by the components of the vector $\a=\{ \a_0, \a_1,..., \a_k \}$.
It's worth pointing out that the parameters $\a_j$, $j=0...k$  do not depend on the specific individual $i$.

This parametrization of $h_i$ correspond to what economists call a
\emph{discrete choice} model \cite{mcfadden}, and shows a
remarkable link between econometrics and statistical mechanics,
which is of special interest in view of McFadden's work concerning
this theory and its application to the study of urban transport.
\newline
Discrete choice theory states that, when making a choice, each
person weights out various factors such as his own gender, age, income,
etc, as to maximize in probability the benefit arising from
his/her decision. Parameters $\a$ tell us the relative weight
(i.e. the importance) that the various socio-economic factors have
when people are making a decision with respect to our binary
question.
The parameter $\a_0$ does not multiply any specific attribute, and thus
it is a homogeneous influence which is felt by all people
in the same way, regardless of their individual characteristics.
A discrete choice model is considered good when the parametrized
attributes are very suitable for the specific choice, so that
the parameter $\a_0$ is found to be small in comparison to the
attribute-specific ones.

Elementary statistical mechanics tells us that the probability of
an individual $i$ with attributes $a_i$ answering ``YES'' to our
question is the following \cite{ellis}:
\begin{eqnarray*}
p_i&=&P(\s_i=1)=\frac{e^{h_i}}{e^{h_i}+e^{-h_i}}
\\
h_i&=&\sum_{j=1}^k \a_j a_i^{(j)}+\a_0
\end{eqnarray*}
Therefore collecting the choices made by a relevant number of
people and keeping track of their socio-economic attributes
allows us to use a statistics in order to find the value of $\a$
for which our distribution best fits the real data. This in turn
allows to assess the implications on aggregate behavior if we
apply incentives to the population which affect specific
attribute, as can be commodity prices in a market situation.
%

\section{Interaction}

The kind of model described in the last section has been
successfully used by econometrics for the last thirty years
\cite{mcfadden}, and has opened the way to the quantitative study
of social phenomena. Such models, however, only apply to
situations where the functional relation between the people's
attributes $\a$ and the population's behavior is a smooth one: it is
ever more evident, on the other hand, that behavior at a societal
level can be marked by sudden jumps \cite{bouchaud2, salganik,
kuran}.

There exist many examples from linguistics, economics, and sociology
where it has been observed how the global behaviour of
large groups of people can change in an abrupt manner as a
consequence of slight variations in the social structure (such as,
for instance, a change in the pronunciation of a language due to a
little immigration rate, or as a substantial decrease in crime rates
due to seemingly minor action taken by the authorities)
\cite{critmass, gladwell, kuran}. From a statistical mechanical
point of view, these abrupt transition may be considered
as phase transitions caused by the interaction between
individuals. Indeed, Brock and Durlauf have shown
\cite{durlauf} how discrete choice can be extended to the case
where a global mean-field interaction is present (providing
an interesting mapping to the Curie-Weiss theory \cite{ellis}), thus
further highlighting the close relation existing between the
econometric and the statistical mechanical approaches to problems
concerning many agents.

We then go back to studying the general interacting model
\begin{equation}\label{ham}
H_N(\s)=-\sum_{i,l=1}^N J_{il}\s_i\s_l - \sum_{i=1}^N h_i \s_i,
\end{equation}
while keeping
$$
h_i=\sum_{j=1}^k \a_j a_i^{(j)}+\a_0.
$$

We now need to find a suitable parametrization for the interaction coefficients $J_{il}$. Since each person is
characterized by $k$ binary socio-economic attributes,
the population can be naturally partitioned into $2^k$ subgroups, which for convenience we take of
equal size: this leads us to consider a mean field
kind of interaction, where coefficients $J_{il}$ depend explicitly on such a partition. We can express this as
follows
\begin{eqnarray*}
J_{il}=\frac{1}{2^k \, N}J_{gg'}, \ \textrm{if $i \in g$ and $l \in g'$},
\end{eqnarray*}
which in turn allows us to rewrite (\ref{ham}) as
\begin{eqnarray*}\label{intens}
H_N(\s)=-\frac{N}{2^k}(\sum_{g,g'=1}^{2^k} J_{gg'}m_g m_{g'} +
\sum_{g=1}^{2^k} h_g m_g)
\end{eqnarray*}
where $m_g$ is the average opinion of group $g$:
$$
m_g=\frac{1}{2^k \, N}\sum_{i=(g-1)N/2^k+1}^{g \, N/2^k} \s_i.
$$

In \cite{gaco} the case $k=1$ of this model was considered: the model's thermodynamic limit was proved, and it was
given a rigorous derivation of the model's solution, as well as an analysis of some analytic properties.
In particular, it was shown that the model factorizes completely, so that all the information
about the model consists of the equilibrium states:
\begin{eqnarray}\label{mfe}
\bar m_1 &=& \tanh(J_{11} \bar m_1 + J_{12} \bar m_2 + h_1)
\\
\bar m_2 &=& \tanh(J_{21} \bar m_1 + J_{22} \bar m_2 + h_2)
\end{eqnarray}

This allows us, in particular, to write the probability of $i$ choosing YES in a closed form,
similar to the non-interacting one:
\begin{equation}\label{prob}
p_i=P(\s_i=1)=\frac{e^{U_g}}{e^{U_g}+e^{-U_g}},
\end{equation}
where
$$
U_g=\sum_{g'=1}^2 J_{g,g'} \bar m_{g'} + h_g.
$$

This is the basic tool needed to estimate the model starting from real data. We describe
the estimation procedure in the next section.

\section{Estimation}

We have seen that according to the model an individual $i$ belonging to group $g$ has
probability of choosing ``YES'' equal to
$$
p_i=\frac{e^{U_g}}{e^{U_g}+e^{-U_g}}
$$
where
$$
U_g=\sum_{g'} J_{g,g'} \bar m_{g'} + h_g.
$$

The standard approach of statistical estimation for discrete
models is to maximize the probability of observing a sample of
data with respect to the parameters of the model (see
e.g. \cite{benakiva}). This is done by maximizing the likelihood
function \cite{greene}
$$
L=\prod_i p_i
$$
with respect to the model's parameters, which in our case consist of the interaction
matrix $J$ and the vector $\alpha$.

Our model, however, is such that $p_i$ is a function of the equilibrium states $m_g$,
which in turn are {\it discontinuous} functions of the model's parameters.
This problem takes away much of the appeal of the maximum likelihood procedure,
and calls for a more feasible alternative.

The natural alternative to maximum likelihood for problems of
model regression is given by the least squares method
\cite{greene}, which simply minimizes the squared norm of the
difference between observed quantities, and the model's
prediction. Since in our case the observed quantities are the
empirical average opinions $\tm_g$, we need to find the parameter values
which minimize
\begin{equation}\label{least}
    \sum_g (\tm_g - \tanh U_g)^2,
\end{equation}
which in our case correspond to satisfying as closely as possible
the state equations (\ref{mfe}) in squared norm. This, however, is
still computationally cumbersome due to the non-linearity of the
function $\tanh(U_j)$. This problem has already been encountered
by  Berkson back in the nineteen-fifties, when developing a
statistical methodology for bioassay \cite{berkson}: this is an
interesting point, since this stimulus-response kind of experiment
bears a close analogy  to the natural kind of
applications for a model of social behavior, such as linking
stimula given by incentive through policy and media, to behavioral
responses on part of a population. Furthermore the same approach is used by
statistical mechanics, for example within the
problem of finding  the proper order parameter for a given
Hamiltonian \cite{barra0}.

The key observation in Berkson's paper is that, since $U_g$ is a linear function of the model's parameters,
and the function $\tanh(x)$ is invertible, a viable modification to least squares is given by minimizing the
following quantity, instead:
\begin{equation}\label{berk}
    \sum_g (\arctanh \tm_g -  U_g)^2.
\end{equation}
This reduces the problem to a linear least squares problem which can be handled with standard statistical
software, and Berkson finds an excellent numerical agreement between this method and the standard least squares procedure.

There are nevertheless a number of issues with Berkson's approach,
which are analyzed in \cite{benakiva}, pag. 96. All the problems
arising can be traced to the fact that to build (\ref{berk}), we
are collecting the individual observations into subgroups, each of
average opinion $m_g$. The problem is well exemplified by the case
in which a subgroup has average opinion $m_g \equiv \pm 1$: in
this case $\arctanh m_g=-\infty$, and the method breaks down.
However the event $m_g \equiv \pm 1$ has a vanishing probability
when the size of the groups increases, so that the method behaves
properly for large enough samples.
\newline
The proposed measurement technique is best elucidated by showing a
few simple concrete examples, which we do in the next section.

\section{Case studies}

We shall carry out the estimation program for real situations which
corresponds to a very simple case of our model.
The data was obtained from periodical censuses carried out by Istat  \footnote{Italian National Institute of Statistics}:
since census data concerns events which are recorded in official documents, for a large number of people, we find it
to be an ideal testing ground for our model.
\newline
For the sake
of simplicity, individuals  are described  by a single binary attribute characterizing their place of residence
(either Northern or Southern Italy) and we chose, among the several possible case studies, the ones for which choices are likely to involve peer interaction in a  major way.
\newline
The first phenomenon we choose to study concerns the share of people who
chose to marry through a religious ritual, rather than through a
civil one. The second case deals with divorces:  here individuals are faced with the choice
 of a consensual/ non-consensual divorce.
The last test we perform regards the study of suicidal tendencies, in particular the mode of execution.
\newline

\subsection{Civil vs religious marriage in Italy, 2000-2006}

To address this first  task  we use data from the annual report on
the institution of marriage compiled by Istat in the seven years going from
$2000$ to $2006$.
\newline
The reason for choosing this specific social question is both a methodological and a conceptual one.
\newline
Firstly, we are motivated by the exceptional quality of the data
available in this case, since it is a census which concerns a
population of more  than $250$ thousand people per year, for seven
years. This allows us some leeway from the possible issues
regarding the sample size, such as the one highlighted in the last
section. And just as importantly the availability of a time series
of data measured at even times also allows to check the
consistency of the data as well as the stability of the
phenomenon.

Secondly, marriage is probably one of the few matters where a great number of individuals
makes a genuine choice concerning their life that gets recorded in an official document, as opposed
to what happens, for example, in the case of opinion polls.

We choose to study the data with one of the simplest forms of the model: individuals are divided
according to only to a binary attribute $a^{(1)}$, which takes value $1$ for people from Northern
Italy, and $0$ for people form Southern Italy.
In the formalism of Section $2$, therefore, the model is defined
by the Hamiltonian
\begin{eqnarray*}
H_N(\s) &=& - \frac{N}{2} ( J_{11} m_1^2 + (J_{12}+J_{21}) m_1 m_2 + J_{22} m_2^2 +  h_1 m_1 +  h_2 m_2),
\\
h_i     &=& \a_1 a_i^{(1)}+\a_0,
\end{eqnarray*}
and the state equations to be used for Berkson's statistical procedure are given by (\ref{prob}).
\begin{center}
\begin{table}
\begin{tabular}{r|c|c|c|c|c|c|c|}
 & \multicolumn{7}{|c|}{{\textbf \% of religious marriages, by year}} \\
\hline
{\textbf Region} & {\textbf 2000} & {\textbf 2001} & {\textbf 2002} & {\textbf 2003} & {\textbf 2004} & {\textbf 2005} & {\textbf 2006} \\
\hline
{\textbf Northern Italy} & 68.35 & 64.98 & 61.97 & 54.64 & 57.91 & 55.95 & 54.64 \\
{\textbf Southern Italy} & 81.83 & 80.08 & 79.32 & 75.46 & 76.81 & 76.52 & 75.46 \\
\hline
\end{tabular}
\caption{Percentage of religious marriages, by year and geographical region}\label{ts1}
\end{table}
\end{center}
Table \ref{ts1} shows the time evolution of the share of men
choosing to marry through a religious ritual: the population is
divided in two geographical classes. The first thing worth noticing is that
these shares show a remarkable stability over the seven-year
period: this confirms how, though arising  from choices made by
distinct individuals, who bear extremely different personal
motivations, the aggregate behavior can be seen as an observable
feature characterizing society as a whole.


In order to apply Berkson's method of estimation, we choose gather
the data into periods of four years, starting with $2000-2003$,
then $2001-2004$, etc. Now, if we label the share of men in group
$g$ choosing the religious ritual in a specific year (say in
$2000$) by $m_g^{2000}$, we have that the quantity that ought to
be minimized in order to estimate the model's parameters for the
first period is the following, which we label $X^2$:
\begin{eqnarray*}
   X^2 &=& \sum_{year=2000}^{2003} \sum_{g=1}^2 (\arctanh m^{year}_g -  U_g^{year})^2,
\\
    U_g^{year}&=&\sum_{g'=1} ^2J_{g,g'} m_{g'}^{year} + h_g,
\\
    h_g     &=& \a_1 a_g^{(1)}+\a_0.
\end{eqnarray*}

The results of the estimation for the four periods are shown in Table \ref{J1}, whereas Table \ref{H1} shows the corresponding
estimation for a discrete choice model which doesn't take into account interaction.
\begin{center}
\begin{table}
\begin{tabular}{r|ccc|ccc|ccc|ccc|}
 & \multicolumn{12}{|c|}{{ 4-year period}} \\
\hline
{Parameter} & \multicolumn{3}{|c|}{{\textbf 2000-2003}} & \multicolumn{3}{|c|}{{\textbf 2001-2004}} & \multicolumn{3}{|c|}{{\textbf 2002-2005}} & \multicolumn{3}{|c|}{{\textbf 2003-2006}} \\
\hline
\hline
{$\a_0$} & -0.10 & $\pm$ & 0.42 & -0.16 & $\pm$ & 0.15 & -0.18 & $\pm$ & 0.10 & -0.13 & $\pm$ & 0.01 \\
{$\a_1$} & 0.20 & $\pm$ & 0.59 & 0.20 & $\pm$ & 0.22 & 0.16 & $\pm$ & 0.14 & 0.14 & $\pm$ & 0.01 \\
{$J_1$} & 1.16 & $\pm$ & 0.41 & 1.09 & $\pm$ & 0.16 & 1.01 & $\pm$ & 0.11 & 1.02 & $\pm$ & 0.01 \\
{$J_2$} & 1.29 & $\pm$ & 0.89 & 1.40 & $\pm$ & 0.33 & 1.45 & $\pm$ & 0.21 & 1.36 & $\pm$ & 0.01 \\
{$J_{12}$} & -0.21 & $\pm$ & 0.89 & -0.10 & $\pm$ & 0.33 & 0.03 & $\pm$ & 0.21 & -0.01 & $\pm$ & 0.01 \\
{$J_{21}$} & 0.09 & $\pm$ & 0.41 & 0.02 & $\pm$ & 0.16 & -0.01 & $\pm$ & 0.11 & 0.01 & $\pm$ & 0.01 \\
\hline
\end{tabular}
\caption{Religious vs civil marriages: estimation for the interacting model}\label{J1}
\end{table}
\begin{table}
\begin{tabular}{r|ccc|ccc|ccc|ccc|}
 & \multicolumn{12}{|c|}{{\textbf 4-year period}} \\
\hline
{\textbf Parameter} & \multicolumn{3}{|c|}{{\textbf 2000-2003}} & \multicolumn{3}{|c|}{{\textbf 2001-2004}} & \multicolumn{3}{|c|}{{\textbf 2002-2005}} & \multicolumn{3}{|c|}{{\textbf 2003-2006}} \\
\hline
{$\a_0$} & 0.67 & $\pm$ & 0.15 & 0.63 & $\pm$ & 0.03 & 0.61 & $\pm$ & 0.06 & 0.58 & $\pm$ & 0.03 \\
{$\a_1$} & -0.41 & $\pm$ & 0.1 & -0.43 & $\pm$ & 0.04 & -0.45 & $\pm$ & 0.08 & -0.46 & $\pm$ & 0.04 \\
\hline
\end{tabular}
\caption{Religious vs civil marriages: estimation for the non-interacting model}\label{H1}
\end{table}
\end{center}

\subsection{Divorces in Italy, 2000-2005}

The second case study uses data from the annual report compiled by Istat in the six years
going from $2000$ to $2005$. The data show how divorcing couples chose between a consensual and a non-consensual
 divorce in Northern  and Southern Italy.
\newline
As shown in Table \ref{ts2} here too, when looking at the ratio among consensual versus the total divorces, the data show a remarkable stability.
\newline
Again we gather the data into periods of four years and Table \ref{J2} presents the estimation of our model's parameters for the whole available period, while in Table \ref{H2} we show the corresponding fit by the non-interacting discrete choice model.

We notice that the  estimated parameters have some analogies with the preceding case study in that here too the cross interactions $J_{12}, \ J_{21}$ are statistically close to zero whereas the diagonal values $J_{11}, \ J_{22}$ are both greater than one suggesting an interaction scenario which is due to multiple equilibria \cite{gaco}. Furthermore, in both cases the attribute-specific parameter $\a_1$
is larger than the generic parameter $\a_0$ in the interacting model (Tables 2 and 5), as opposed to what we see in the
non-interacting case
(Tables 3 and 6): this suggests that by accounting for interaction we might be able to better evaluate the role played by
socio-economic attributes.
\begin{center}
\begin{table}
\begin{tabular}{r|r|r|r|r|r|r|}
& \multicolumn{6}{|c|}{{\textbf \% of consensual divorces, by year}} \\
\hline
{\textbf Region} & {\textbf 2000} & {\textbf 2001} & {\textbf 2002} & {\textbf 2003} & {\textbf 2004} & {\textbf 2005} \\
\hline
{\textbf Northern Italy} & 75.06 & 80.75 & 81.32 & 81.62 & 81.55 & 81.58 \\
{\textbf Southern Italy} & 58.83 & 72.80 & 71.80 & 72.61 & 72.76 & 72.08 \\
\hline
\end{tabular}
\caption{Percentage of consensual divorces, by year and geographical region}\label{ts2}
\end{table}
\end{center}

\begin{center}
\begin{table}
\begin{tabular}{r|ccc|ccc|ccc|}
 & \multicolumn{9}{|c|}{{\textbf 4-year period}} \\
\hline
{\textbf Parameter} & \multicolumn{3}{|c|}{{\textbf 2000-2003}} & \multicolumn{3}{|c|}{{\textbf 2001-2004}} & \multicolumn{3}{|c|}{{\textbf 2002-2005}} \\
\hline
\hline
{$\a_0$} & 0.02 & $\pm$ & 0.06 & -0.08 & $\pm$ & 0.01 & -0.07 & $\pm$ & 0.01 \\
{$\a_1$} & -0.25 & $\pm$ & 0.08 & -0.22 & $\pm$ & 0.01 & -0.23 & $\pm$ & 0.01 \\
{$J_1$} & 1.59 & $\pm$ & 0.14 & 1.64 & $\pm$ & 0.01 & 1.66 & $\pm$ & 0.01 \\
{$J_2$} & 1.16 & $\pm$ & 0.06 & 1.25 & $\pm$ & 0.01 & 1.25 & $\pm$ & 0.01 \\
{$J_{12}$} & -0.05 & $\pm$ & 0.06 & 0.01 & $\pm$ & 0.01 & 0.00 & $\pm$ & 0.01 \\
{$J_{21}$} & -0.08 & $\pm$ & 0.14 & 0.00 & $\pm$ & 0.01 & -0.01 & $\pm$ & 0.01 \\
\hline
\end{tabular}
\caption{Consensual vs non-consensual divorces: estimation for the interacting model}\label{J2}
\begin{tabular}{r|ccc|ccc|ccc|}
 & \multicolumn{9}{|c|}{{\textbf 4-year period}} \\
\hline
{\textbf Parameter} & \multicolumn{3}{|c|}{{\textbf 2000-2003}} & \multicolumn{3}{|c|}{{\textbf 2001-2004}} & \multicolumn{3}{|c|}{{\textbf 2002-2005}} \\
\hline
{$\a_0$} & 0.41 & $\pm$ & 0.13 & 0.48 & $\pm$ & 0.01 & 0.480046 & $\pm$ & 0.01 \\
{$\a_1$} & 0.28 & $\pm$ & 0.18 & 0.25 & $\pm$ & 0.02 & 0.261956 & $\pm$ & 0.01 \\
\hline
\end{tabular}
\caption{Consensual vs non-consensual divorces: estimation for the non-interacting model}\label{H2}
\end{table}
\end{center}

\subsection{Suicidal tendencies in Italy, 2000-2007}

The last case study deals with suicidal tendencies in Italy, again following the  annual report compiled by Istat in the six years  from $2000$ to $2007$, and we use the same geographical attribute used for the former two studies.

The data in Table \ref{ts3} shows the percentage of deaths due to hanging as a mode of execution. The topic of suicide is
of particular relevance to sociology: indeed, the very first systematic quantitative treatise in the social sciences was
carried out by \'Emile Durkheim \cite{durkheim}, a founding father of the subject, who was puzzled by how a phenomenon as unnatural as
suicide could arise with the astonishing regularity that he found. Such a regularity as even been dimmed the ``sociology's one law'' \cite{pope}, and there is hope that the connection to statistical mechanics might eventually shed light on the origin of such a law.

Mirroring the two previous case studies, we present the time series in Table \ref{ts3}, whereas Table \ref{J3} shows the
estimation results for the interacting model, and Table \ref{H3} are the estimation results for the discrete choice model.
Again, the data agrees with the analogies found for the two previous case studies. \newline

\begin{center}
\begin{table}
\begin{tabular}{r|c|c|c|c|c|c|c|c|}
 & \multicolumn{8}{|c|}{{\textbf \% of suicides by hanging}} \\
\hline
{\textbf Region} & {\textbf 2000} & {\textbf 2001} & {\textbf 2002} & {\textbf 2003} & {\textbf 2004} & {\textbf 2005} & {\textbf 2006} & {\textbf 2007} \\
\hline
{\textbf Northern Italy} & 34.17 & 37.02 & 35.83 & 34.58 & 35.21 & 36.23 & 33.57 & 38.08 \\
{\textbf Southern Italy} & 37.10 & 37.40 & 37.34 & 38.54 & 34.71 & 38.90 & 40.63 & 36.66 \\
\hline
\end{tabular}
\caption{Percentage of suicides with hanging as mode of execution, by year and geographical region}\label{ts3}
\end{table}
\begin{table}
\begin{tabular}{r|c|r|r|r|r|}
 & \multicolumn{5}{|c|}{{\textbf 4-year period}} \\
\hline
{\textbf Parameter} & \multicolumn{1}{|c|}{{\textbf 2000-2003}} & \multicolumn{1}{|c|}{{\textbf 2001-2004}} & \multicolumn{1}{|c|}{{\textbf 2002-2005}} & \multicolumn{1}{|c|}{{\textbf 2003-2006}} & \multicolumn{1}{|c|}{{\textbf 2004-2007}} \\
\hline
\hline
{$\a_0$} & 0.01  $\pm$  0 & 0.02  $\pm$  0.01 & 0.01 $\pm$  0.01 & 0.02  $\pm$  0.01 & 0.02  $\pm$ 0.01 \\
{$\a_1$} & 0.01  $\pm$  0.01 & 0.00  $\pm$  0.01 & 0.00  $\pm$  0.01 & 0.00  $\pm$  0.01 & 0.00  $\pm$  0.01 \\
{$J_1$} & 1.09  $\pm$ 0.01 & 1.09  $\pm$  0.01 & 1.09  $\pm$  0.02 & 1.10  $\pm$  0.03 & 1.09  $\pm$  0.01 \\
{$J_2$} & 1.06  $\pm$  0.01 & 1.08  $\pm$ 0.01 & 1.08 $\pm$ 0.01 & 1.07 $\pm$  0.01 & 1.07  $\pm$  0.01 \\
{$J_{12}$} & 0  $\pm$ 0.01 & 0.00  $\pm$  0.01 & 0.00  $\pm$  0.01 & 0.00  $\pm$ 0.01 & 0.00  $\pm$  0.01 \\
{$J_{21}$} & 0  $\pm$  0.01 & 0.01  $\pm$  0.01 & 0.00  $\pm$  0.02 & 0.01  $\pm$ 0.03 & 0.01  $\pm$  0.01 \\
\hline
\end{tabular}
\caption{Suicidal tendencies: estimation for the interacting model}\label{J3}
\end{table}
\end{center}
\begin{center}
\begin{table}
\begin{tabular}{r|c|c|c|c|c|}
 & \multicolumn{5}{|c|}{{\textbf 4-year period}} \\
\hline
{\textbf Parameter} & \multicolumn{1}{|c|}{{\textbf 2000-2003}} & \multicolumn{1}{|c|}{{\textbf 2001-2004}} & \multicolumn{1}{|c|}{{\textbf 2002-2006}} & \multicolumn{1}{|c|}{{\textbf 2003-2007}} & \multicolumn{1}{|c|}{{\textbf 2004-2008}} \\
\hline
{$\a_0$} & -0.25  $\pm$  0.02 & -0.27 $\pm$  0.03 & -0.26  $\pm$  0.03 & -0.24 $\pm$  0.04 & -0.25 $\pm$  0.05 \\
{$\a_1$} & -0.05  $\pm$  0.03 & -0.03  $\pm$ 0.04 & -0.04  $\pm$  0.04 & -0.07 $\pm$  0.06 & -0.04 $\pm$  0.07 \\
\hline
\end{tabular}
\caption{Suicidal tendencies: estimation for the non-interacting model}\label{H3}
\end{table}
\end{center}

\section{Comments}
%
We introduced a class of simple mean field models of choice in the
presence of social interaction, which generalizes
the model introduced in \cite{durlauf}.
\newline
After showing how our model reduces to a standard discrete choice model
when we neglect interaction, we analyzed the simplest kind of interaction
(by accounting for only one attribute): in this case the model reduces to
a well known bipartite model, whose thermodynamic limit as well as
multiple equilibria have already been shown to exist \cite{gaco}.
\newline
In order to test our model we considered three case studies, concerning
relevant social phenomena such as marriage, divorce, and suicide, and we
found that Berkson's method of estimation \cite{berkson} provides
a valuable statistical tool, alternative to the more typical maximum likelihood procedure
used in econometrics, which is not suitable for our model due to discontinuities
arising in its probability structure.
\newline
This papers aims to suggest the outline of a method that can be used to study more specific
situations, where individuals may be modelled in a more precise way, by assigning
more socio-economic attributes to them. In this simple case we were able to find
 consistencies in the interaction parameters regarding different topics for the
 same population.  Furthermore, the parameters values where found to be in a regime
 characterized by multiple equilibria, which suggests the possibility that a
 refinement of this study will eventually lead to the capability of predicting
 abrupt transitions at a societal level.

\section*{Acknowledgment}
The authors thank Elena Agliari, Raffaella Burioni, Cristian Giardina and Laurie Davies for
useful conversations. A.B. and I.G.  acknowledge support by
the CULTAPTATION project (European Commission contract FP6-2004-
NEST-PATH-043434). AB thanks support by the MiUR within the SMARTLIFE
Project (Ministry Decree 13/03/2007 n.368) too. P.C. thanks the CULTAPTATION project
and the Strategic Research Program of University of Bologna for partial financial support.


\begin{thebibliography}{100}

\bibitem{critmass} Ball P., {\em Critical Mass}, {\it Arrow books}: United kingdom,
(2004)

\bibitem{barra0} A. Barra,
\textit{The mean field Ising model trhought interpolating
techniques}, J. Stat. Phys.\textbf{145}, 234-261, (2008)


\bibitem{benakiva} Ben-Akiva M., Lerman S. R., {\em Discrete Choice Analysis},
{\it The MIT Press}, Cambridge, Mass: (1985)

\bibitem{berkson} Berkson J., {\em A statistically precise and relatively simple method of estimating the bioassay with quantal
                            response, based on the logistic function}, {\it Journal of the American Statistical Association}
                            {\bf 48}: 565-599, (1953)

\bibitem{bouchaud1} Borghesi C., Bouchaud J. P.,  {\em Of songs and men: a model for multiple choice with herding},
{\it Quality and Quantity}, {\bf 41}: 557-568, (2007)

\bibitem{durlauf} Brock W., Durlauf S., {\em Discrete Choice with Social Interactions},
{\it Review of Economic Studies}, {\bf 68}: 235-260, (2001)

\bibitem{chandler} Chandler D., \textit{Introduction to modern statistical mechanics}, Oxford University Press, (1987).

\bibitem{durkheim} Durkheim E., \textit{Le Suicide. \'Etude de sociologie}, Paris Alcan, (1897).


\bibitem{lowe} Cont R., L¨owe M., {\em Social distance, heterogeneity and social interaction},
{\it Centre des Math´ematiques Appliqu´ees, Ecole Polytechnique},
R.I. No 505: (2003)

\bibitem{ellis} R.S. Ellis,
\textit{Large deviations and statistical mechanics}, Springer, New
York (1985).

\bibitem{follmer} F\"ollmer H., {\em Random Economies wih Many Interacting Agents}, {J. Math. Econ.}, {\bf 1}: 51-62,
(1973)

\bibitem{greene} D. A. Hensher, J. M. Rose, W. H. Greene \textit{Applied Choice Analysis: A Primer}, Massachusetts, Cambridge University Press, (2005).


\bibitem{galam} Galam S., Moscovici S., {\em Towards a theory of collective phenomena:
Consensus and attitude changes in groups}, {European Journal of Social
Psychology}, {\bf 21}: 49-74, (1991)


\bibitem{gaco} Gallo I., Contucci P., {\em Bipartite Mean Field Spin Systems. Existence and Solution},
{\it MPEJ}, {\bf 14}, (2008)

\bibitem{gacocoga} Gallo F., Contucci P., Coutts A., Gallo I., {\em Tackling climate change through energy efficiency:
                mathematical models for evidence-based public policy recommendations}, {\it arXiv:0804.3319},
                (2008)


\bibitem{gladwell} Gladwell M., {\em The Tipping Point}, {\it Little, Brown and Company},
(2000)


\bibitem{granovetter} Granovetter M., {\em Threshold models of collective behaviour}, {Am. J. Sociol.}, {\bf 83}: 1420-1443,
(1978)

\bibitem{halpern} Halpern D., Bates C., Mulgan G., Aldridge S., {\em Personal Responsibility and Changing
            Behaviour: the state of knowledge and its implications for public policy},
                    {\it  Cabinet Office, S Unit, G Britain},        (2004)

\bibitem{nesh} B. Kenneth  \textit{Game Theory and The Social Contract}. MIT Press (1998).

\bibitem{kuran} Kuran T., {\em Now Out of Never}, {\it World politics},
 (1991)

\bibitem{luce} Luce, R.,  {\em Individual choice behavior: a theoretical analysis}, J.Wiley
and Sons, New York (1959)

\bibitem{mcfadden} Mcfadden D., Economic Choices,
{\it The American Economic Review}, {\bf 91}: 351-378, (2001)

\bibitem{bouchaud2} Michard Q., Bouchaud J. P., {\em Theory of collective opinion shifts: from smooth trends to abrupt swings},
{\it The European Physical Journal B}, {\bf 47}: 151-159, (2001)

\bibitem{pope} Pope W., Danigelis N., {\em Sociology's One Law},
{\it Social Forces}: (1981)



\bibitem{salganik} Salganik M. J., Dodds P. S., Watts D. J., {\em Experimental Study of Inequality and
                    Unpredictability in an Artificial Cultural Market},
                     {\it Science}, {\bf 311}: (2006)

\bibitem{schelling} Schelling T., {\em Micromotives and Macrobehaviour}, { \it W W Norton \& Co Ltd}:
(1978)

\bibitem{soet} Soetevent A. R., Kooreman P., {\em A discrete-choice model with social interactions: with
                    an application to high school teen behavior},
                    {\it Journal of Applied Econometrics}, {\bf 22}: 599-624,
                    (2007)

\bibitem{watts} Watts D. J., Dodds P. S., Influentials, Networks and Public Opinion Formation,
                    {\it Journal of Consumer Research}, {\bf 34}: 441-458,
                    (2007)



\end{thebibliography}
\end{document}